\documentclass[fleqn,usenatbib]{mnras}

\usepackage{newtxtext,newtxmath}

\usepackage[T1]{fontenc}
\usepackage{ae,aecompl}

\usepackage{graphicx}	
\usepackage{amsmath}	
\usepackage{xcolor}

\newcommand{\angstrom}{\text{\normalfont\AA}}

\newcommand{\red}[1]{{\color{black} #1}}
\newcommand{\referee}[1]{{\color{black} #1}}

\title[Magnetic Fields \& Atmospheric Escape]{The Effects of Magnetic Fields on Observational Signatures of Atmospheric Escape in Exoplanets: Double Tail Structures}

\author[S. Carolan et al.]{
S. Carolan,$^{1}$\thanks{E-mail: carolast@tcd.ie}
A. A. Vidotto,$^{1,2}$
G. Hazra$^{1,2}$
C. Villarreal D'Angelo,$^{3}$
D. Kubyshkina$^{1}$
\\
$^{1}$School of Physics, Trinity College Dublin, College Green, Dublin 2, Ireland\\
$^{2}$Leiden Observatory, Leiden University, PO Box 9513, 2300 RA, Leiden, The Netherlands\\
$^{3}$Instituto de Astronomía Téorica y Experimental (CONICET-UNC). Laprida 854, X5000BGR. C\'ordoba, Argentina
}

\date{Accepted XXX. Received YYY; in original form ZZZ}

\pubyear{2021}

\begin{document}
\label{firstpage}
\pagerange{\pageref{firstpage}--\pageref{lastpage}}
\maketitle

\begin{abstract}
Using 3D radiative MHD simulations and Lyman-$\alpha$ transit calculations, we investigate the effect of magnetic fields on the observational signatures of atmospheric escape in exoplanets. Using the same stellar wind, we vary the planet's dipole field strength ($B_p$) from 0 to 10G. For $B_p<3$G, the structure of the escaping atmosphere begins to break away from a comet-like tail following the planet ($B_p=0$), as we see more absorbing material above and below the orbital plane. For $B_p\geq3$G, we find a ``dead-zone'' around the equator, where low velocity material is trapped in the closed magnetic field lines. The dead-zone separates two polar outflows where absorbing material escapes along open field lines, leading to a double tail structure, above and below the orbital plane. We demonstrate that atmospheric escape in magnetised planets occurs through polar outflows, as opposed to the predominantly night-side escape in non-magnetised models. We find a small increase in escape rate with $B_p$, though this should not affect the timescale of atmospheric loss. As the size of the dead-zone increases with $B_p$, so does the line centre absorption in Lyman-$\alpha$, as more low-velocity neutral hydrogen covers the stellar disc during transit. For $B_p<3$G the absorption in the blue wing decreases, as the escaping atmosphere is less funnelled along the line of sight by the stellar wind. In the red wing (and for $B_p>3$G in the blue wing) the absorption increases caused by the growing volume of the magnetosphere. Finally we show that transits below and above the mid-disc differ caused by the asymmetry of the double tail structure.

\end{abstract}




\section{Introduction}

Exoplanets that orbit close to their host stars can lose their atmospheres through photoevaporation, as they receive large amounts of high-energy flux at these orbits \citep{Lammer2003,Baraffe2004,Yelle2004}. The lifetime of the planet's atmosphere depends strongly on the rate of this escape \citep[e.g.,][]{Johnstone2015,Daria2020}. This has a number of interesting implications for the exoplanet community. While of course being crucial for understanding planetary habitability \citep{Lingam2018, Dong2018}, atmospheric escape is also thought to shape the distribution of observed exoplanets \citep{Kurokawa2014, Owen2018, Berger2020}. The so called ``radius gap", an under-population of short period exoplanets with radii between 1.5 and 2 Earth radii \citep{Beaug2013, Fulton2017}, has been attributed to atmospheric loss. The mechanism of this loss is not certain, with photoevaporation and core-powered mass loss both being likely candidates, with both probably contributing over the planet's evolution \citep{Rogers2021}. Direct observations of atmospheric escape have also been found. During transit the escaping atmosphere causes excess absorption, as absorbing elements and molecules cover a larger area on the stellar disc. The ever growing list of these observations contains hot Jupiters such as HD209458b \citep{VM2003} and HD189733b \citep{LDE2010, LDE2012, Jensen2012, BJB2013}, as well as warm Neptunes like GJ436b \citep{Kulow2014, Ehrenreich2015} and GJ3470b \citep{Bourrier2018}. 

To fully understand these observations, we need 3D simulations capable of modelling the escaping atmosphere. One very important factor that must be considered is the stellar wind \citep{McCann2019, Carolan2021, Carolina2021}. As the escaping atmosphere leaves the planet it is soon met by the stellar wind, which shapes this outflow into structures like a comet-like tail following the planet in its orbit. The geometry and extent of these structures around the planet is sensitive to several external parameters such as the ram pressure of the stellar wind, as well as the orbital velocity and tidal forces from the star, which arise due to these close-in orbits \citep{Matsakos2015, Pillitteri2015, Shaikhislamov2016}. The stellar wind can also affect the atmospheric escape rate, whereby a strong stellar wind can prevent the dayside escaping atmosphere from reaching super-sonic velocities \citep{Vidotto2020}. As a result the inner most regions of the escaping atmosphere can be affected, and the escape rate reduced due to a lack of dayside acceleration \citep{Christie2016, Vidotto2020, Carolan2020, Carolan2021}. As this interaction is largely asymmetric, 3D models are required to fully accurately model atmospheric escape. Recently there has been an increasing number of 3D works which include the stellar wind \citep[][\red{Hazra et al., submitted, Kubyshkina et al., in prep}]{Bisikalo2013, Shaikhislamov2016, Schneiter2016, CN2017, Carolina2018, McCann2019, Khodachenko2019, Esquivel2019, Debrecht2020, Carolina2021, Carolan2020, Carolan2021, MacLeod2021}, however the effect of magnetic fields on this interaction, the escape rate and observational signatures of this escape is not yet fully understood.

There have been a number of works that investigate different aspects of how the planet's magnetic field affects escape and its observational signatures. Using 2D models, \citet{Trammell2014} showed that the transit depth increases strongly with magnetic field strength when the hydrogen ionisation layer is magnetically dominated, while in the same year \citet{Owen2014} demonstrated that the strength and geometry of the stellar magnetic field is crucial to determine the fraction of open field lines around the planet. \citet{Khod2015} found that the atmospheric escape rate is weakly affected by field strengths <0.3G, but reduced by an order of magnitude for a 1G field, which could potentially have huge implications on the lifetime of the planet's atmosphere. Though these have all used 2D models, \citet{Matsakos2015} performed 3D simulations of close-in magnetic star-planet interactions, identifying 4 classifications: bowshock, colliding winds, strong planetary wind causing accretion, and Roche-lobe overflow. \citet{Arakcheev2017} found a 70\% reduction in WASP-12's escape rate with a model containing the planet's magnetic field and stellar wind, though not the magnetic field of the stellar wind / star. Using 3D global simulations of HD 209458b, \citet{Carolina2018} showed that the shape of the Lyman-$\alpha$ line depends on both the stellar and planetary fields, as they control the geometry of the magnetosphere and the amount of neutrals inside it. Finally \citet{Harbach2021} used MHD models to demonstrate the dependence of the magnetosphere and Lyman-$\alpha$ absorption on the stellar wind, though their model did not self-consistently calculate the heating and photoionisation due to stellar UV irradiation, and ignored close-in orbital forces. To the best of our knowledge, to date there has not been a 3D self-consistent radiative magnetohydrodynamic model capable of resolving the inner most regions of the escaping atmosphere, which contains both the stellar and planetary magnetic fields. 

In this work we present such a model, and use it to investigate the effect of magnetic fields on the escaping atmosphere and their observational signatures. While keeping all other parameters constant (see table \ref{tab:table}) we vary the dipole strength of the planet's magnetic field from 0 to 10 G, examining the change in the magnetosphere and mid transit absorption in Lyman-$\alpha$. The details of our 3D model are given in section \ref{sec:modelling}. In section \ref{sec:outflow} we discuss the effects of the planetary magnetic field on the geometry of the escaping atmosphere, while in section \ref{sec:observations} we discuss the implications of this on the observational signatures present in Lyman-$\alpha$. A discussion of the impact of our results can be found in section \ref{sec:discussion}, while we sum up our conclusions in section \ref{sec:conclusions}.

\section{3D self-consistent radiative magnetohydrodynamic simulations}
\label{sec:modelling}

\begin{figure}
    \centering
    \includegraphics[width=\columnwidth]{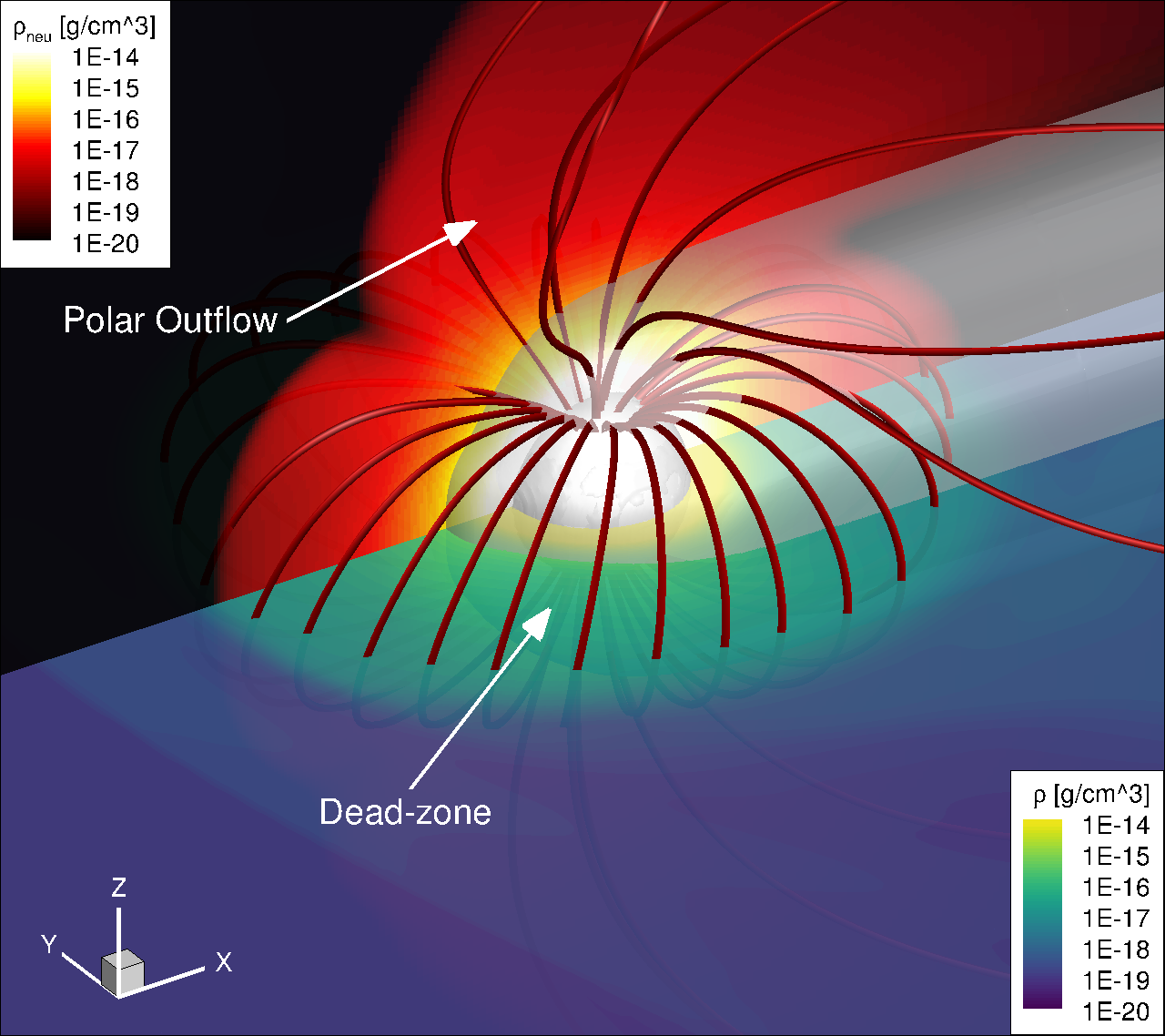}
    \caption{A 3D view of the 10G model. The white sphere is the planet, while the grey surface mark where the optical depth $\tau=1$ of Lyman-$\alpha$ photons. The black-red contour shows the density of neutrals on the polar plane, while the blue-green contour shows the total mass density on the orbital plane. The red lines trace the planetary magnetic field lines. Here we can clearly see a dead-zone of material trapped by the closed field lines, as well as a polar outflow where the field lines are open.}
    \label{fig:3dfigure}
\end{figure}

To model the escaping atmosphere in the presence of a magnetic field, we build upon the non-magnetised code presented in \red{Hazra et al (submitted)}, which uses the BATS-R-US framework \citep{Toth-swmf}. This model is an extensive update to \citet{Carolan2020, Carolan2021}'s model, including new physics: the heating, cooling, ionisation and recombination of neutral and ionised hydrogen, calculated during runtime.  
While  the simulations of \red{Hazra et al. (submitted)} were hydrodynamic, we adapt this model to investigate how magnetic fields affect atmospheric escape and its observational signatures. This is the first 3D self-consistent radiative magnetohydrodynamic  model of photoevaporation of an exoplanet using the BATS-R-US framework. The output of one such model is shown in figure \ref{fig:3dfigure}.

We simulate the escaping atmosphere in a 3D Cartesian grid [-30:50, -40:40, -30:30] planetary radii ($R_p$) with a maximum resolution of 1/16 $R_p$ inside 5 $R_p$. For simplicity we assume that the exoplanet is centered on the origin of the coordinate system, is tidally locked to its host star located outside our numerical domain at negative x, and the simulation is performed in the co-rotating, co-orbiting reference frame. The model solves the ideal magnetohydrodynamic equations: the mass conservation; momentum conservation; energy conservation and induction equations respectively:

\begin{equation}
\frac{\partial{\rho}}{\partial{t}} +\nabla \cdot \rho {\vec{ u}} = 0,
\end{equation}
\begin{equation}\label{eqn:momentum}
\frac{\partial(\rho\vec{u})}{\partial t} + \nabla \cdot \Bigg[\rho \Vec{u} \Vec{u} + (P_T + \frac{B^2}{8\pi})I - \frac{\Vec{B}\Vec{B}}{4\pi}\Bigg] = \rho\bigg( \vec{g} -\frac{GM_{*}}{|\vec{R}|^2} \hat{R} - \vec{\Omega} \times (\vec{\Omega}\times\vec{R})-2(\vec{\Omega} \times \vec{u}) \bigg), 
\end{equation}
%
\begin{eqnarray}\label{eqn:energy}
\frac{\partial \epsilon}{\partial t} + \nabla \cdot \Bigg[\Vec{u}\Bigg(\epsilon + P_T + \frac{B^2}{8\pi}\Bigg)-\frac{(\Vec{u}\cdot\Vec{B})\Vec{B}}{4\pi}\Bigg] = ~\nonumber \\ \rho \bigg( \vec{g} -\frac{GM_{*}}{|\vec{R}|^2} \hat{R} - \vec{\Omega} \times (\vec{\Omega}\times\vec{R}) \bigg) \cdot \vec{u} 
+{\cal H}-{\cal C},
\end{eqnarray}
%
\begin{equation}
    \frac{\partial\Vec{B}}{\partial t} + \nabla \cdot (\Vec{u}\Vec{B} - \Vec{B}\Vec{u}) = 0.
\end{equation}

\noindent $\rho$, $\vec{u}$, $P_T$, $\vec{B}$ and $I$ are the mass density, velocity, thermal pressure, magnetic field, and identity matrix respectively. \referee{$\vec{R}$ is the position vector relative to the centre of the star, given by $\vec{R} = \vec{r} + \vec{a}$, where $\vec{r}$ is the position vector relative to the centre of the planet, and $\vec{a}$ is the orbital distance}. The total energy density  $\epsilon$ is 
\begin{equation}
    \epsilon = \frac{\rho u^2}{2} + \frac{P_T}{\gamma -1} + \frac{B^2}{8\pi}, 
\end{equation}

\noindent $\gamma$ is the adiabatic index, which we set to 5/3. In the momentum equation \ref{eqn:momentum}, the source terms in order are the planet's gravity, the stellar gravity, the centrifugal and Coriolis forces. The energy conservation equation \ref{eqn:energy} contains ${\cal H}$ and ${\cal C}$ terms, denoting the volumetric heating and cooling rates. The volumetric heating rate due to stellar radiation is given by: 

\begin{equation}
   {\cal H} = \eta \sigma n_n F_{\rm xuv}e^{-\tau},  
\end{equation}

\noindent here $\eta$ is the excess energy released when a hydrogen atom is ionised, $n_n$ is the number density of neutrals, $F_{\rm xuv}$ is the incident XUV flux and $\tau$ is the optical depth. We assume that the incident XUV radiation is plane parallel, and that the entire XUV spectrum is concentrated  at 20 eV. This yields $\sigma$= $1.89\times10^{-18}$ cm$^{-2}$  and $\eta$= 0.32 \citep{MurrayClay2009,Allan2019,Hazra2020} \red{Hazra et al. (submitted)}. The XUV flux is injected into the grid from the negative x boundary, such that the optical depth is calculated by:

\begin{equation}
    \tau = \int_{x_{\rm boundary}}^{x}n_n \sigma dx.
\end{equation}

For the total volumetric cooling rate ${\cal C}$, our model contains the cooling due to Lyman-$\alpha$ emission \citep{Osterbrock1989}:

\begin{equation}
{\cal C}_{{\rm Ly}\alpha}= 7.5 \times 10^{-19} n_p n_n \exp{(-1.183 \times 10^5/T )},
\end{equation} 
and the cooling due to collisions \citep{Black1981}:
\begin{equation}
 {\cal C}_{\rm col} = 5.83 \times 10^{-11} n_e n_n   \sqrt{T} \exp{({-1.578\times 10^5}/{T})} \chi_H,
\end{equation}
where T is the temperature, $\chi_H = 2.18\times 10^{-11} $erg is the ionisation potential of hydrogen, and $n_p$ and $n_e$ are the number density of  protons and electrons in cm$^{-3}$. This yields volumetric heating and cooling rates in units of erg cm$^{-3}$ s$^{-1}$.

In addition to these, our model solves two additional mass conservation equations, tracking the density of neutrals and ions:

\begin{equation}
   \frac{\partial n_n}{\partial t} + \nabla \cdot n_n\vec{u} = \mathscr{R} - \mathscr{I},
\end{equation}
\begin{equation}
   \frac{\partial n_p}{\partial t} + \nabla \cdot n_p\vec{u} = \mathscr{I} - \mathscr{R},
\end{equation}

\noindent where $\mathscr{R}$ and and $\mathscr{I}$ are the the recombination rate and ionisation rate due to photoionisation and collisional ionisation \citep{Osterbrock1989} given by:
\begin{equation}
 \mathscr{R} = 2.7 \times 10^{-13} (10^4/T)^{0.9} n_e n_p,
\end{equation}
\begin{equation}
 \mathscr{I} =\frac{ \sigma n_n F_{\rm xuv} e^{-\tau}}{{h\nu}} + 5.83 \times 10^{-11} n_e n_n \sqrt{T} \exp{({-1.578\times 10^5}/{T})}.
\end{equation}
\noindent where $\mathscr{R}$ and $\mathscr{I}$ are in cm$^{-3}$ s$^{-1}$.

We impose an inner boundary at $1R_p$. Here we fix the base temperature and density to 1000 K and $2.4\times10^{11}$/cm$^3$ respectively. Similar to the 1D models of \citet{MurrayClay2009} we find that changing these values has no significant effect on the escape rate. For velocity, we use a reflective boundary \referee{(the velocity in the true and ghost cells have the same magnitude but the opposite sign)}, such that the velocity of material starts at $\approx$ 0 km/s at $1R_p$. \referee{In spite of the very small initial velocity, the planetary atmosphere is accelerated  above the boundary according to the  forces in our momentum equation. For the magnetic field we use BATS-R-US' Global Magnetosphere module's default boundary \citep[e.g.,][]{DeZeeuw2004, Toth-swmf}}. This fixes the field strengths at $R=0.5R_p$ such that the desired dipole strength is obtained at $R=1R_p$, where a floating boundary condition is applied \referee{(the gradient of magnetic field is kept constant between true and ghost cells, such that the field lines can respond to changes in the outflow)}. For the outer boundaries (with the exception of the negative x boundary for the stellar wind) we use inflow limiting boundary conditions \citep{McCann2019, Carolan2020, Carolan2021}. These are required when simulating in the co-rotating frame to remove any unwanted and uncontrolled inflows associated with the Coriolis force bending material near a boundary. 

\referee{We initialise our computational domain with a 1D $\beta$ profile of the escaping atmosphere, which is fit to a 1D model from \citet{Allan2019}. In the case studied here, this profile takes the form $u_r = u_\infty (1-1/r)^\beta$ where $u_\infty=38$ km/s is the terminal velocity of the outflow, and $\beta = 2.97$ is found as the best fit to the 1D model. From mass conservation, we then initialise the density in the whole grid as $n(r) = n_0 u_0 / u_r$, where $n_0$ is the density at the boundary (see previous paragraph) and $u_0$ is derived from our 1D model. Initially, we assume a constant ionisation fraction throughout the grid (0.001\%), but as the solution advances, the ionisation fraction is self-consistently obtained through Equations (10) and (11). This setup ensures that there is absorbing material (i.e., neutral hydrogen) in the grid to absorb the ionising radiation from the star when the simulation begins. We note that the exact setup of the $\beta$ profile does not affect the resulting steady-state solution. We begin our simulations with just planetary material, and once the escaping atmosphere has reached steady-state, we then turn on the orbital forces and inject the stellar wind, yielding the resulting steady-state solutions seen in figures \ref{fig:orbital_plane} and \ref{fig:polar_plane} respectively.}

We inject the stellar wind at the negative x boundary. Here we use similar boundary conditions to \citet{Carolan2020, Carolan2021}, now adapted to also handle the stellar wind's magnetic field. \referee{The boundary assumes a stellar wind velocity and magnetic field which is radial away from the star.} We provide values of the stellar wind velocity, temperature, density and magnetic field which are derived from an external model, which describes the stellar wind using a 1D polytropic model.
We use a similar model to that of \citet{Carolan2019}, which is based on \citet{johnstone2015a}'s version of VAC \citep{Toth1996}. In polytropic wind models the density and pressure are related by the polytropic index ($\alpha$) according to $P_{sw} \propto \rho_{sw}^\alpha$. We adopt a polytropic index of $\alpha=1.05$, which implies that the temperature of the stellar wind is nearly isothermal. We fix the stellar wind temperature and mass loss rate at the boundary of our 3D simulations, the latter of which is used with the velocity solution from the 1D model to set the density along the boundary ($\rho_{SW} = \dot{M}/4\pi R^2 \vec{u_{SW}} $). \referee{Note that the stellar wind model is separate to our 3D simulations, and is not updated during runtime.} For the purpose of this work, we use the same stellar wind in each simulation (see table \ref{tab:table} for details). The stellar wind is chosen as it is super-alfvénic at the planet's orbit, so  that the interaction with the escaping atmosphere cannot travel upstream in the stellar wind and affect the boundary condition. This will allow us to more consistently examine the effect of magnetic fields on the planetary outflow, as varying the planet's magnetic field cannot affect the injected stellar wind, ensuring the stellar wind is identical in all models. The model parameters for the planet and stellar wind can be found in table \ref{tab:table}.

\begin{table*}
    \centering
    \caption{The parameters of our models. $M_p$ and $R_p$ are the mass and radius of the planet, while $B_{0p}$ is the range of polar dipole field strength of the planet examined. $a$ is the orbital distance from the star. $F_{xuv}$ is the X-ray flux received by the planet. $M_*$ and $R_*$ are the mass and radius of the star. $R_A$ is the alfvén  point of the stellar wind while $\dot{M}_*$ is the stellar wind mass loss rate. $u_{r*}$ is the range of stellar wind velocities along the negative x boundary, with the minimum at the centre and increasing towards the edge of the grid-face. Finally $B_{0*}$ is the stellar dipole strength.
}
    \begin{tabular}{ccccccccccc}
        \hline
        $M_p$ & $R_p$ & $B_{0p}$& $a$ & $F_{xuv}$ & $M_*$ & $R_*$ & $R_A$ & $\dot{M}_*$ & $u_{r*}$ & $B_{0*}$\\
        $[M_J]$ & $[R_J]$ & [G]&  [au] & [erg/cm$^2$/s] & [$M_\odot$] & [$R_\odot$] & [$R_*$] & [$M_\odot/yr$] & [km/s] & [G]\\
        \hline
        0.7 & 1.4 & 0-10&  0.05 & 850 & 1 & 1 & 5.56 & $3.6\times 10^{-13}$ & 327-368 & 2G\\
        \hline
    \end{tabular}
    \label{tab:table}
\end{table*}

\section{Effect of magnetic fields: atmospheric escape}
\label{sec:outflow}
We run 5 models in total using the parameters in table \ref{tab:table}, varying the magnetic field strength of the planet's dipole from 0 to 10G, while keeping the stellar wind constant.

\begin{figure*}
    \centering
    \vspace{-1cm}
    \includegraphics[width=\textwidth]{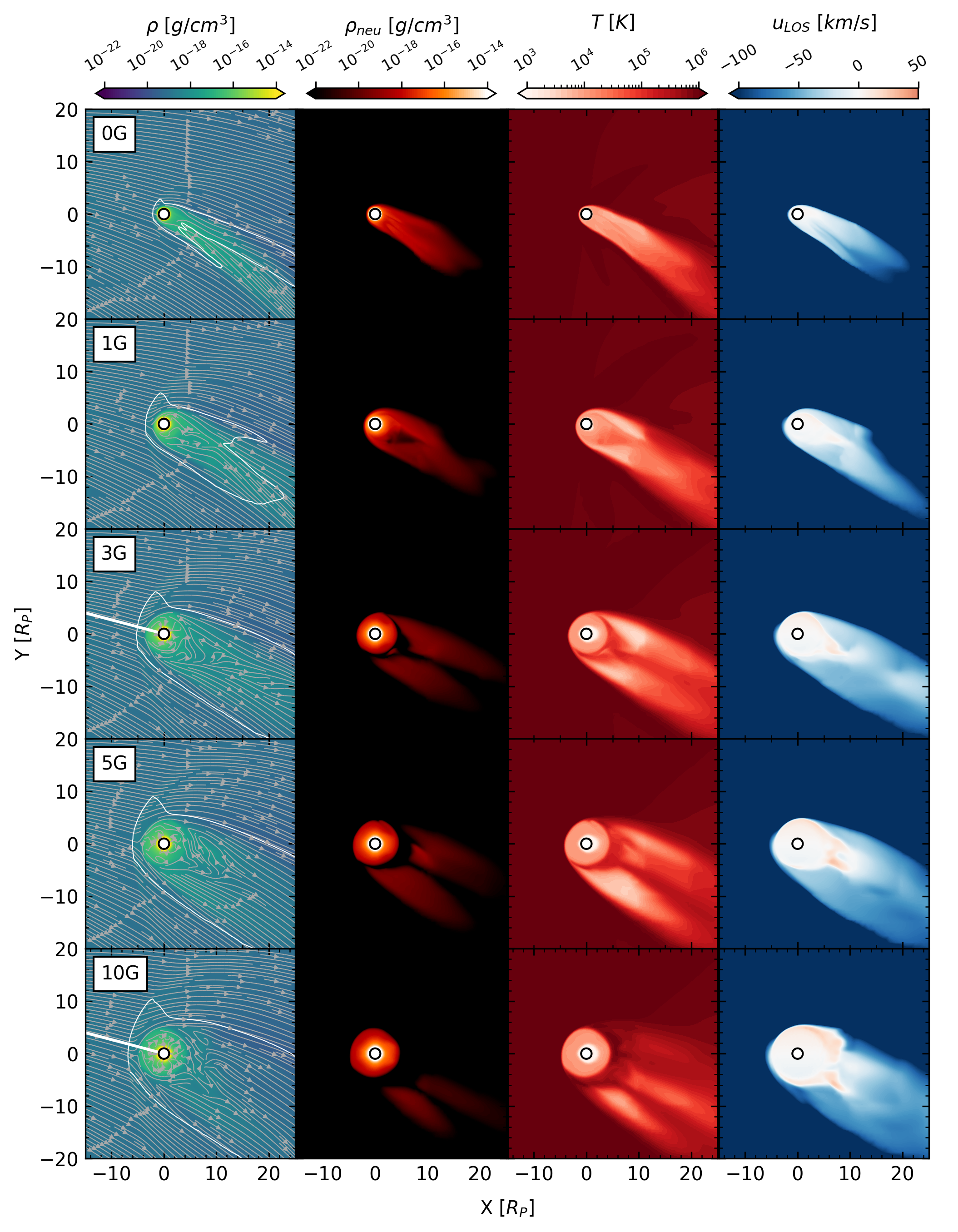}
    \caption{Cut at orbital plane of each of our models. Each row shows a different magnetic field strength of the planet, while each column shows the total density, neutral density, temperature and line of sight velocity respectively. The star is located at negative x, while the planet is marked by a circle centred on the origin. The grey streamlines in the left column trace the flow of material in each model (see figure \ref{fig:polar_plane} for stream-tracers of magnetic field). The white contour in the left column shows the position of the magnetosonic surface. The white line in the left middle panel is the line on which we examine different pressures in figure \ref{fig:pressures}. }
    \label{fig:orbital_plane}
\end{figure*}

\begin{figure*}
    \centering
    \includegraphics[width=\textwidth]{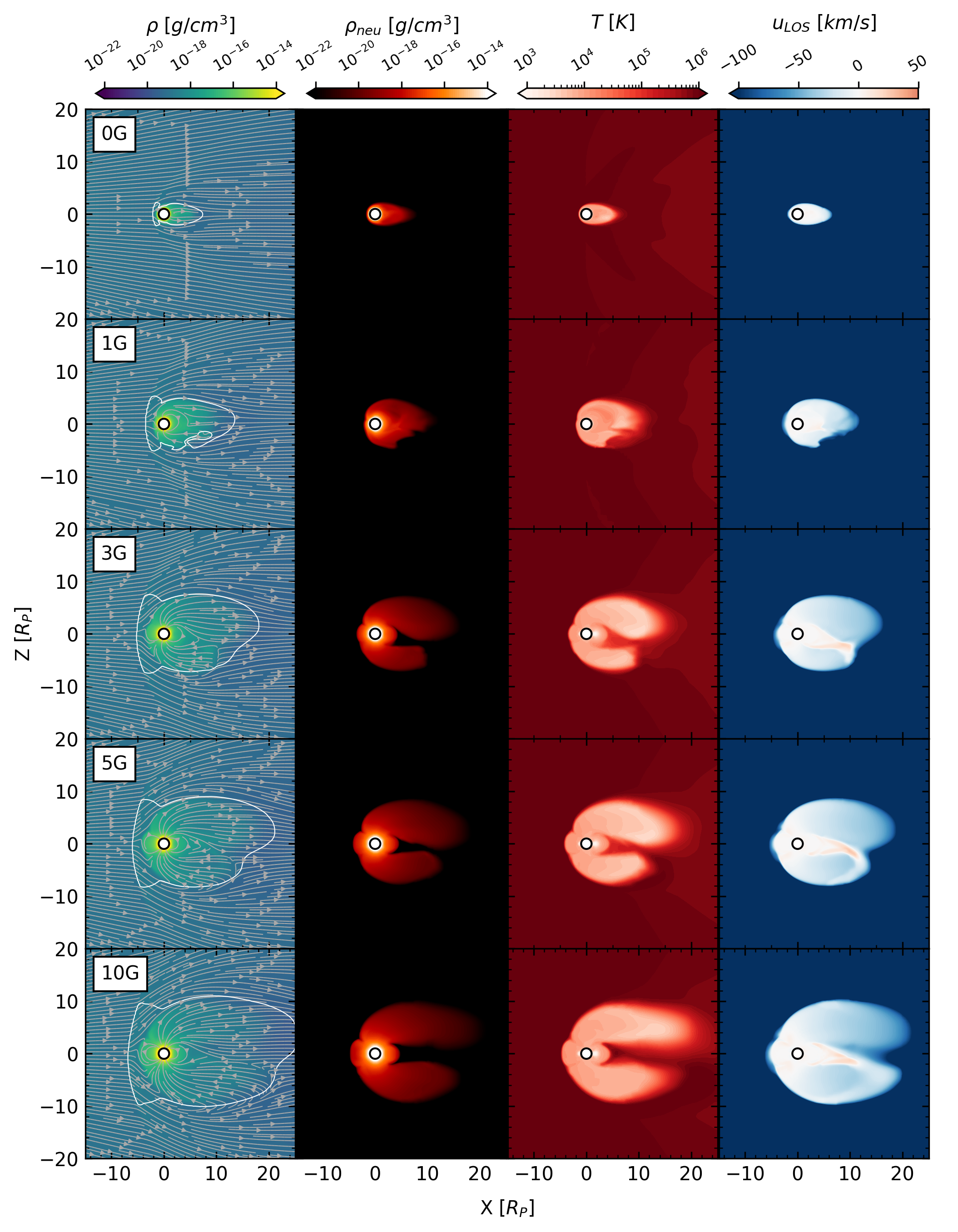}
    \caption{Cut at the polar plane in each of our models, similar to figure \ref{fig:orbital_plane}. The grey streamlines in the left column now trace the magnetic field lines in each model. }
    \label{fig:polar_plane}
\end{figure*}

Figures \ref{fig:orbital_plane} and \ref{fig:polar_plane} show the orbital and polar planes in each of our models, at steady-state (snapshot of quasi steady-state 1G model, see appendix \ref{sec:appendix}). The 0G model resembles closely the strong stellar wind models from \citet{Carolan2020, Carolan2021}, where the stellar wind funnels material into a tight comet-like tail centred on the orbital plane.  We can see that this tail contains all of the low temperature neutral material, most of which has an line of sight velocity that will contribute to blue shifted absorption in Lyman-$\alpha$ (see section \ref{sec:observations}).

The presence of a planetary magnetic field leads to a different tail structure. Similar to other works \citep[eg.][]{Khod2015}, as the planetary magnetic field strength is increased, the structure moves away from an tail centred on the orbital plane. Where the magnetic field lines are open, we obtain two polar outflows, one above and below the orbital plane (as seen in the lower panels of figure \ref{fig:polar_plane}). These are separated by a ``dead-zone'' around the planet where the field lines are closed, filled with mostly low velocity neutral material as seen in figure \ref{fig:3dfigure}. In figure \ref{fig:polar_plane} we see that the size of the dead-zone increases with magnetic field strength, as a growing circle of low temperature and low velocity material is held around the planet on the orbital plane. In figures \ref{fig:orbital_plane} and \ref{fig:polar_plane} we can see that as the size of the dead-zone increases the separation between the polar outflows also increases, until no singular comet-like tail is seen in the orbital plane of the 10G model. Instead we see two tail-like structures following the planet, one above and one below the orbital plane. Note that these two flows are asymmetric which will be discussed in section \ref{sec:asymmetry}. 

For lower magnetic field strengths \red{(<3G)}, most of the absorbing material is at low to blue-shifted velocities. For stronger magnetic fields\red{($\geq$3G)}, we see a small amount of red-shifted material on the night-side of the planet between the polar flows, as the magnetic field funnels some material back towards the planet.
 
\begin{figure}
    \centering
    \includegraphics[width=\columnwidth]{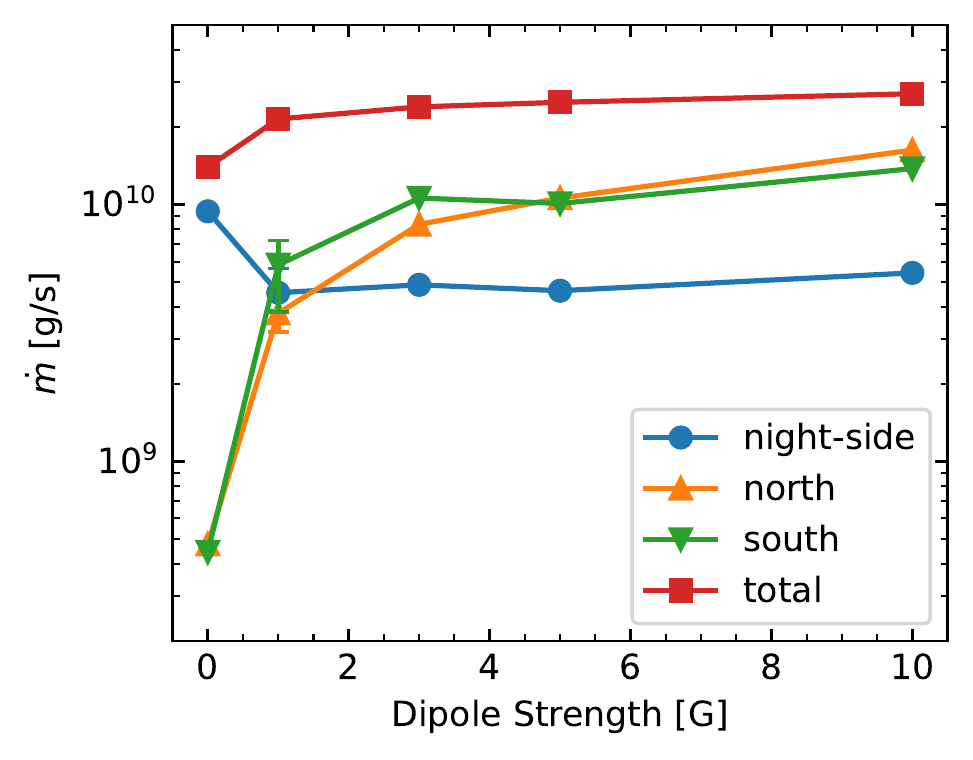}
    \caption{The mass loss rates of each of our models. The blue circles mark the mass lost through a plane at $x=2.5 R_p$  representing the night-side mass loss. The orange and green triangles mark the mass lost through planes at $z=2.5 R_p$ and $z=-2.5   R_p$ respectively, representing the polar flows. The red squares mark the total mass lost, marking the average mass lost through concentric spheres around the planet. \referee{For the 1G model, we plot the mean and variation of the mass-loss rates during the quasi-steady state solution, discussed in appendix \ref{sec:appendix}. }}
    \label{fig:mass_loss_rates}
\end{figure}

\referee{In our models the $\tau=1$ surface varies from $\approx 1.8 R_p$ in the 10G model, to $\approx 1.2 R_p$ in the 0G model, along the sub-stellar line. The exact values for the position of the $\tau=1$ surface in each of our models can be found in table \ref{tab:magnetopause}. As previously discussed, when the magnetic field strength of the planet is increased, the size of the deadzone grows. This larger, denser deadzone is then able to absorb more of the incident stellar radiation before it gets lower into the planet's atmosphere, and as a result the $\tau=1$ surface is pushed to higher altitudes.}

To quantify the change in the escaping atmosphere in our models, we examine the planetary mass loss through planes parallel to the orbital plane above and below the planet at $z=\pm2.5 R_p$ vs that loss through the night-side between these planes at $x=2.5R_p$  (figure \ref{fig:mass_loss_rates}). We can then compare these to the total mass loss calculated through concentric spheres around the planet to investigate where the planet is losing most of its atmosphere. As expected from figures \ref{fig:orbital_plane} and \ref{fig:polar_plane}, the 0G model loses most of its mass through the night-side, as the stellar wind funnels the escaping atmosphere into a tight comet-like tail centred on the orbital plane. Once a planetary magnetic field is introduced, we begin to see much more mass loss through polar outflows than through the night-side, with these polar flows contributing to the majority of atmospheric escape for $B_p\geq3G$. For the 1G model (and the 3G model to a lesser extent) we can see significant difference between the north and south polar flows, which we attribute to the variation in this quasi-steady state solution \red{(see appendix \ref{sec:appendix})}. We note also the increase in the total atmospheric escape rate ($\dot{m}$) seen from the 0 to 10G models \red{(factor of 2)}. This is caused by the funnelling geometry of the magnetic field which assists in the driving of escape in the polar flows, compared to the predominantly night-side outflow in the 0G model. This is similar to what is seen in stellar wind models with magnetic fields \citep{Vidotto2009, vidotto2014,Reville2015, dualta2019, dualta2021,Kavanagh2019,Kavanagh2021}. However for stronger field strengths (>3G) the magnetic field does not cause significant changes to the escape rate, though it does change the observational signatures of this escape, as discussed in section \ref{sec:observations}. 

\begin{figure*}
    \centering
    \includegraphics[width=0.8\textwidth]{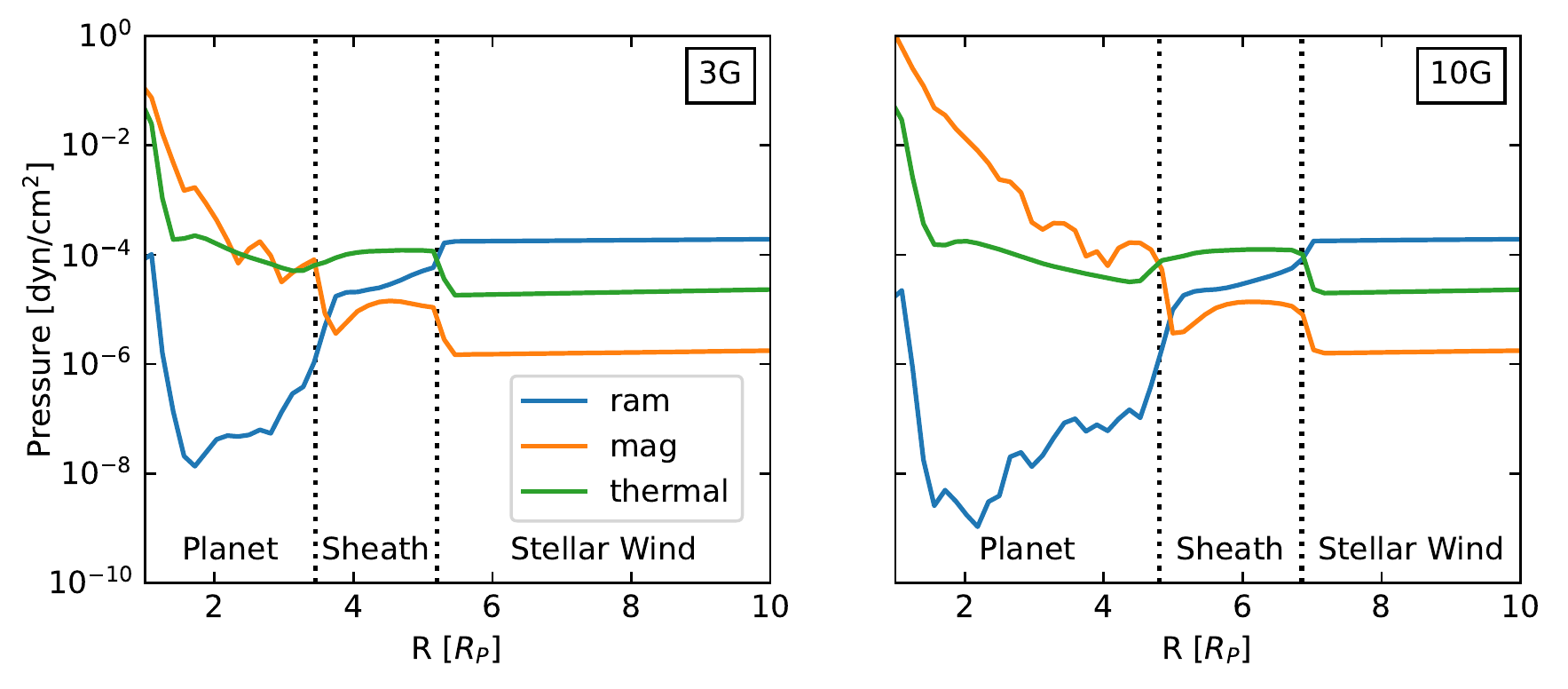}
    \caption{The ram, magnetic and thermal pressures in two of our models, along the lines marked in left middle and bottom panels of figure \ref{fig:orbital_plane}. This line was chosen where the stellar wind velocity is perpendicular to the shock. The black dotted lines mark the position of the magnetopause and bow shock standoff distances, which separate the planetary, magnetosheath, and stellar wind material respectively.}
    \label{fig:pressures}
\end{figure*}

Figure \ref{fig:pressures} shows the ram ($\rho u^2$), magnetic ($B^2/8\pi$) and thermal ($nk_BT$) pressures in our models along the white line shown in figure \ref{fig:orbital_plane}. This is the line where the stellar wind velocity is perpendicular to the bow shock, where the size of the magnetosphere is quantified. As seen in figure \ref{fig:pressures} we find a similar pressure structure to the Earth models of \citet{Carolan2019}, and so we use a similar procedure to quantify the size of the magnetosphere in each of our models, \red{given in table \ref{tab:magnetopause}}. We find that in the stellar wind, ram pressure is the dominant pressure component, while the shocked material in the magnetosheath is dominated by thermal pressure. The bow shock standoff distance can then be found as the point where ram and thermal pressures are balanced. Inside the planetary magnetosphere the dominant pressure component is magnetic, and so the magnetopause is found by the balance of magnetic and thermal pressures. As expected, increasing the magnetic field strength of the planet's dipole increases the size of the magnetosphere, as increasing the magnetic pressure will push the balance of magnetic and thermal pressures further from the planet. Note that due to the lack of intrinsic magnetic field in the 0G model, we cannot identify the magnetopause in this model. In reality it is likely that stellar wind interaction would generate a weak induced magnetic field, which would yield a magnetopause standoff distance $ < 1.5 R_p$. For all models, we observe a ``weak'' bow shock, similar to the 30$\Omega_\odot$ model from \citet{Carolan2019}. This is due to a relatively low mach number of the stellar wind $M_m \approx 1.5$, which causes a jump in density and decrease in velocity by a factor of approx 2.6  \citep[for a strong shock this factor is approximately 4, ][]{Balogh2013, Gombosi2004, Spreiter1966}. We also obtain a relatively thick magnetosheath in our simulations because of this, with the bow shock distance being greater than 1.4 times the magnetopause standoff distance in each model (1.275 in a strong shock).  

\begin{table}
    \caption{The size of the planet's magnetosphere along the line shown in figure \ref{fig:orbital_plane}, chosen as the point where the stellar wind velocity is perpendicular to the shock. $B_p$ is the planet's magnetic field strength in Gauss, $R_m$ and $R_b$ are distance of the magnetopause and bow shock standoff distance relative to the radius of the planet. \referee{$R_\tau$ is the position of the $\tau=1$ surface on the sub-stellar line.}}
    \centering
    \begin{tabular}{cccc}
        \hline
        $B_p$ & $R_m$ & $R_b$ & $R_\tau$\\
        $[G]$ & [$R_p$] & [$R_p$] & [$R_p$]\\
        \hline
        0 & --- & 2.2 & 1.2\\
        1 & 2.1 & 3.5 & 1.4\\
        3 & 3.5 & 5.2 & 1.7\\
        5 & 4.2 & 6.0 & 1.7\\
        10 & 4.8 & 6.9 & 1.8\\
        \hline
    \end{tabular}
    \label{tab:magnetopause}
\end{table}

\section{Effect of magnetic fields: observational signatures}
\label{sec:observations}

To investigate how magnetic fields affect observational signatures of escape, we calculate transit absorption profile of each of our models in Lyman-$\alpha$. To do this we use a ray-tracing method similar to \citet{Carolan2021}. Placing the observer at positive x, we first calculate the line of sight (LOS) velocity. As our 3D model simulates atmospheric escape in the planet's reference frame, we must adjust the velocity in our grid to the line of sight velocity by $u_{LOS} = -u_x + y\Omega$. As our simulations track the density of neutral hydrogen, we can extract the temperature, density of neutrals and line of sight velocity from our grid and begin the transit calculation.

The frequency ($\nu$) dependent optical depth is given by:

\begin{equation}
    \tau_\nu = \int n_n \sigma \phi_\nu dx,
\end{equation}

\noindent where $\sigma = {\pi e^2 f}/{m_e c}$ is the absorption cross section at line centre, and $\phi_\nu$ is the Voigt line profile. $f = 0.416410$ is the oscillator strength for Lyman-$\alpha$, $m_e$ is the mass of the electron, $e$ is the electron charge and $c$ is the speed of light. Once the optical depth is calculated we can then find the fraction of incident specific intensity that is transmitted:

\begin{equation}
    \frac{I_\nu}{I_*} = e^{-\tau_\nu}.
\end{equation}

\noindent $1 - \frac{I_\nu}{I_*}$ is therefore the fraction of specific intensity that is absorbed. We divide our grid into 201x201 columns and calculate this intensity in 51 velocity bins between -500 and 500 km/s. The transit depth can then be found as:

\begin{equation}\label{eqn_depth}
    \Delta F_\nu = \frac{\int\int (1-e^{-\tau_\nu})dydz}{\pi R_*^2}.
\end{equation}

\begin{figure}
    \centering
    \includegraphics[width=\columnwidth]{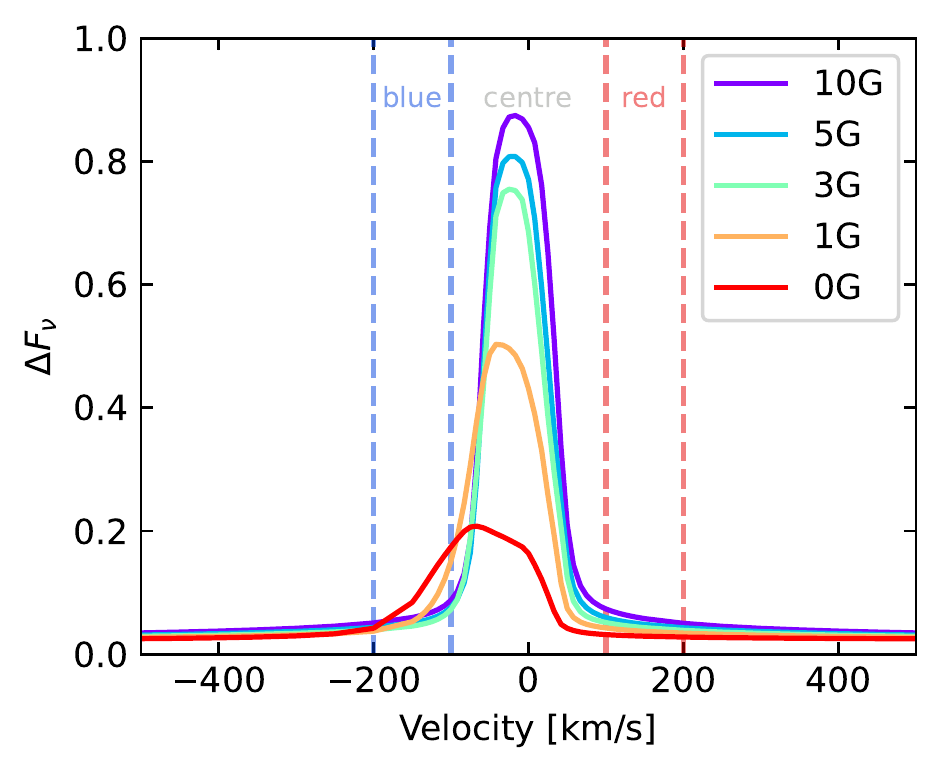}
    \caption{The mid transit absorption profiles for each of our models, for an impact parameter $b=0.$ The blue and red dashed lines mark the bounds for the integrals of the blue, line centre and red wing absorptions in figure \ref{fig:deltaF}}
    \label{fig:line_profiles}
\end{figure}

\noindent The transit depth of each model at mid transit can be seen in figure \ref{fig:line_profiles}.

\red{Even though the variation in $\dot{m}$ is small,} varying the strength of the planet's magnetic field causes significant changes to absorption in Lyman-$\alpha$. As the comet-like tail in the 0G model is tightly funnelled down the line of sight by the stellar wind, we see significant blue wing absorption in this model. For small magnetic field strengths the tail begins to extend further above and below the orbital plane \footnote{This is the case where the planet's dipole axis is perpendicular to the orbital plane. However if the dipole was tilted on the polar plane, the separation between the polar flows and the orbital axis would be reduce see section 5.1}. The magnetic field now introduces a larger obstacle to the stellar wind. As a result the escaping atmosphere is less accelerated along the line of sight,  and so we do not see as much blue-shifted absorption.

We see a significant increase in line centre absorption with increasing magnetic field strength. This is due to the growing dead-zones around the planet. As the planet's dipole strength increases, the size of the dead-zone increases as shown in figures \ref{fig:orbital_plane} and \ref{fig:polar_plane}. This region contains mostly low velocity, low temperature neutrals, and so as the size of the dead-zone grows, so does the line centre absorption.

\subsection{Transit Asymmetry due to Wind-Planet Magnetic Interaction}
\label{sec:asymmetry}

In figure \ref{fig:polar_plane} we see asymmetry in the planetary material above and below the orbital plane. This is due to the interaction between the stellar wind and planetary magnetic fields. When the super-alfvénic stellar wind reaches the planet, it is shocked and deflected around the planet. This deflection is what shapes the comet-like tail following the planet. As the stellar wind is shocked to sub-alfvénic velocities, reconnection of field lines can occur, which we see happen just inside the alfvén surface. Due to the orientation of the planetary magnetic field and radial stellar field in our models, the stellar and planetary field lines reconnect at the south pole, below the orbital plane, while the northern planetary field lines are opened. We note that as the upstream stellar wind is super-alfvénic, this reconnection cannot accelerate particles that travel back to the star (this is believed to cause enhanced activity in the star \citealt[][]{Shkolnik2018,Cauley2019,Folsom2020, Kavanagh2021}, which could change the properties of the stellar wind). As a result of this reconnection the absorbing neutral material is less extended below the orbital plane, as the bending magnetic field lines at the interface between the stellar wind and escaping atmosphere act as a boundary (see figure \ref{fig:polar_plane}). Above the orbital plane (north pole) there are now more open field lines allowing the neutral material to extend further compared below the plane.

\begin{figure}
    \centering
    \includegraphics[width=\columnwidth]{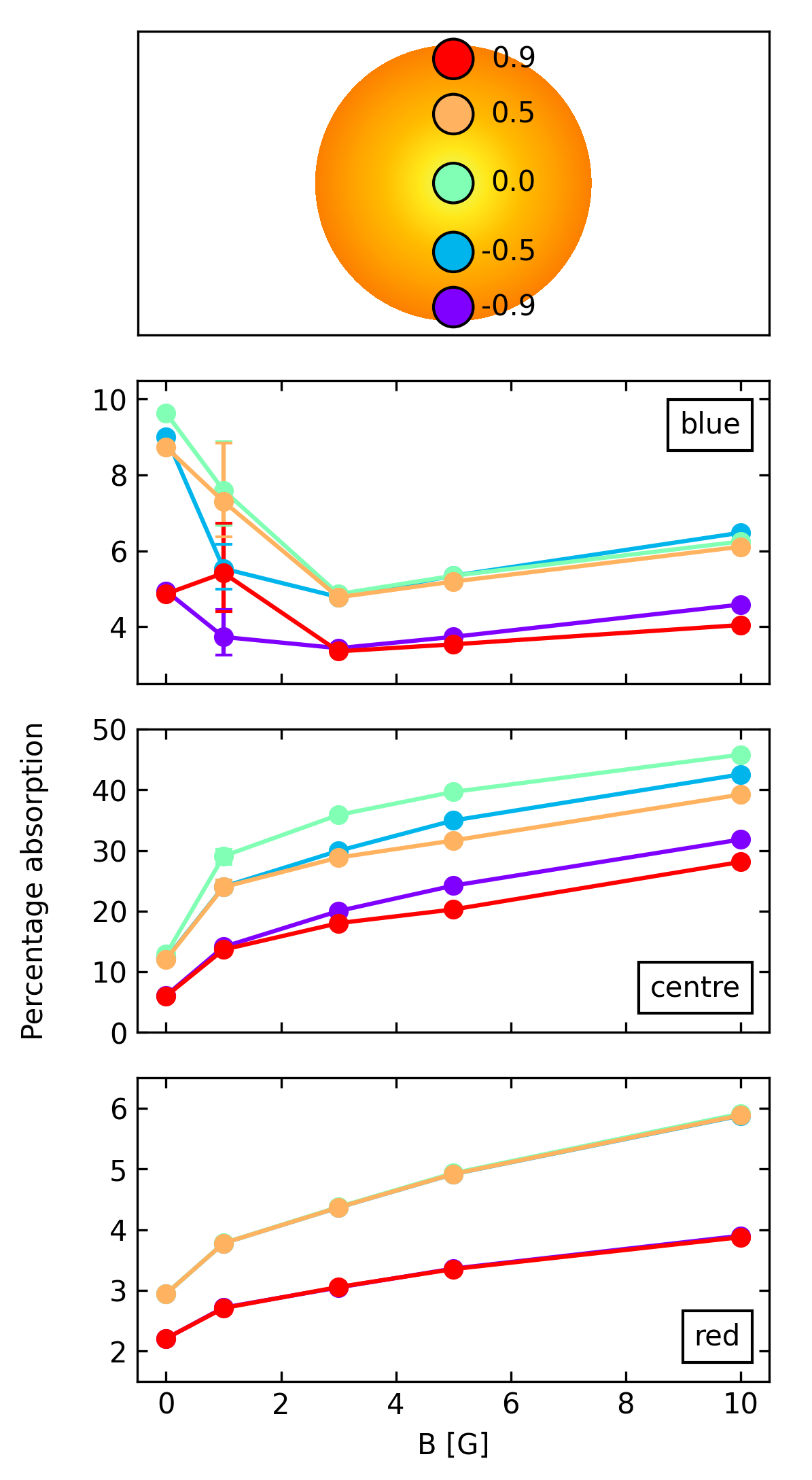}
    \caption{The blue, line centre and red wing absorptions (marked on figure \ref{fig:line_profiles}) of each of our models for a variety of impact parameters. The top panel illustrates the position of the planet at mid transit as it transits horizontally across the stellar disc. \referee{In the second panel we plot the mean and variation in the blue wing absorption for the 1G model during its quasi-steady state solution (see appendix \ref{sec:appendix}). Note that there is no such variation in the red wing, or at line centre.}}
    \label{fig:deltaF}
\end{figure}

Due to the asymmetry in the distribution of absorbing neutrals above and below the orbital plane we can expect this to affect the absorption profiles depending on the transit geometry of the planet. In figure \ref{fig:deltaF} we show the percentage absorptions in the blue, line centre and red wings of the Lyman-$\alpha$ line shown in figure \ref{fig:line_profiles}, for a variety of impact parameters ($b$) in the northern and southern hemispheres of the stellar disc. 

In the line centre, mid disc ($b=0$) has the largest absorption, as it maximises the volume of material within the stellar disc (green curves).
In the blue wing we initially see a decrease in absorption for increasing magnetic field strength. As previously mentioned this is because the structure of the comet-like tail moves away from being funnelled fully along the line of sight on the orbital plane, as the polar outflows perpendicular to the orbital plane are introduced. The most asymmetry between positive and negative impact parameters (see top panel of figure \ref{fig:deltaF}) is seen in the 1G model. As previously mentioned this model reaches a quasi-steady state, showing some variation in the distribution of material. As we do not see this in other simulations with higher $B_p$, we believe this is due to the magnetic field not being able to fully break this single comet-like tail structure into the polar outflows and dead-zones we see in other models with higher $B_p$. \referee{During} the quasi-steady state solution for this model, a positive impact parameter shows more blue wing absorption than the negative values, suggesting that more high velocity Lyman-$\alpha$ absorption is caused by material under the orbital plane. \referee{This is caused be the disparity in the mass lost through the southern pole in this model (see figure \ref{fig:mass_loss_rates}), as the magnetic field begins to break the single comet-like tail structure.} We note that depending on the stage during the quasi-steady state variation examined, the velocity of the material under the plane will vary significantly and so this asymmetry will not remain constant (see appendix \ref{sec:appendix}).

For strong planetary fields, we see that blue wing absorption increases with magnetic field strength. As the magnetic field strength increases the size of the magnetosphere increases, allowing more room for absorbing material to accelerate along the polar outflows. As expected from figure \ref{fig:polar_plane} we see some asymmetry between the positive and negative impact parameters. The negative values yield higher absorption than the positive counterparts. For positive impact parameters the more extended northern flows will lie outside the stellar disc (while it will lie inside the disc for negative impact parameters), yielding less absorption in Lyman-$\alpha$. This is also the case for absorption in line centre. As there are more neutrals above the plane the negative impact parameters will allow more lower velocity absorbing material to cover the disc than positive values.

There is no asymmetry present in red shifted material when comparing positive and negative impact parameters. From figures \ref{fig:orbital_plane} and \ref{fig:polar_plane} we can see that most redshifted material lies on the orbital plane between the polar outflows on the night-side of the planet, as a small amount of material falls from the polar outflows back towards the planet. As a result there is no significant difference between the positive and negative impact transits.

\section{Discussion}
\label{sec:discussion}
\subsection{B-field Geometries}
As the absorption is related on the geometry of the escaping atmosphere, which is dependent on a combination of $B_*$ and $B_p$, our results \red{are expected to} change for different \red{magnetic} field topologies. For example, if $B_*$ had the opposite polarity, then our results for positive and negative impact parameters would be inverted. In this scenario, the reconnection of stellar and planetary field lines now occurs above the orbital plane (as opposed to below, shown in figure \ref{fig:polar_plane}). This would lead to more open field lines below than above the plane. As a result, the distribution of material we currently see in our models would be flipped through the orbital plane, leading to the opposite result for transits with positive and negative impact parameters. Similarly, if the planetary dipole was tilted, the position of reconnection would change. The topology used in this work, where the planetary dipole axis is perpendicular to both the sub-stellar line and direction of transit, yields the greatest difference between positive and negative impact parameter transits. \red{Tilting the dipole towards / away from the star will reduce the vertical distance between the two polar outflows, as rotating the dipole brings the poles closer to the orbital plane. As a result of this reduced separation, the observed asymmetry between transit geometry would also be reduced.}  Similarly tilting the dipole towards/away from the transit direction (towards positive y ) will reduce asymmetry. Transits close to the edge of the stellar disc now no longer fully remove one of the polar outflows, but rather parts of both, producing less asymmetry between transit geometries.

\subsection{No Significant Change in Escape Rate}
In this work we find that increasing the planet's magnetic field strength causes a minor increase in the planet's total mass loss rate (increasing by a factor of 2). This result is contrary to what other works have found for hot Jupiters \citep[eg.][ all found reductions by up to an order of magnitude]{Owen2014,Khod2015, Arakcheev2017}, and likely caused by difference in the physics included in each model. Similar to other works, we find that the inclusion of magnetic fields creates dead-zones in our models around the equator, where planetary material is held inside the closed field lines. This reduces the mass loss around the equator, while also causing an increase in the line centre absorption as these dead-zones hold a significant amount of low velocity neutrals. Despite this, all of our models with planetary magnetic fields lose more mass than our 0G model. As previously discussed, in these models more mass is funnelled through the polar flows, leading to two outflows above and below the equatorial dead-zones. Unlike \citet{Khod2015}, we do not find suppressed polar winds, where faster adiabatic cooling leads to reduced polar winds. This could be in part due to the inclusion of stellar wind in our model. While \citet{Khod2015} found most mass is loss through the ``wind zone'' between the equatorial dead-zones and suppressed polar winds, the inclusion of the stellar wind in our model deflects material from the dayside ``wind zone'' towards the night-side comet-like tails. Through this deflection material lost through the dayside ``wind zone'' must flow through where the suppressed polar outflows would reside to reach the comet-like tail, leading to no apparent suppression of polar winds in our models. Instead we see an increase in polar outflows when compared to the 0G model. This increased polar flow accompanied by the decreased night-side mass loss nets an increase in the total mass loss rate by a factor of 2, though this is not significant enough to affect the timescale for atmospheric loss. This further emphasises the importance of considering the stellar wind when modelling atmospheric loss, as it can significantly alter the the geometry and rate of atmospheric escape.

\subsection{Implications on Observations}
The results we present in this work have important implications on interpreting transit observations. Despite the mass-loss rate only increasing by a factor of 2, increasing the magnetic field strength greatly alters the absorption line profile. At line centre, the total percentage absorption increases by a factor of 4 at mid transit, when comparing the 0 and 10G models. The presence of a magnetic field greatly alters the geometry and distribution of absorbing material in the planetary magnetosphere, and so is crucial to consider in order to correctly interpret transit observations.
One issue with observing Lyman-$\alpha$ transits is that this line is not observable at line centre, due to both interstellar absorption and geocoronal emission. Recent work has identified the infra-red $10830\angstrom$ Helium I triplet as an atmospheric escape identifier \citep[eg.][]{Spake2018, Nortmann2018}. A popular approach has been to calculate the population of the triplet in post processing, i.e., a Parker-type wind is used to calculate the bulk properties of the escaping atmosphere and, afterwards, the population state is calculated \citep[eg.][Dos Santos et al. submitted]{Oklopcic2018, Lampon2020, MacLeod2021}. There have also been numerous detections of heavier elements in the transmission spectra of exoplanets \citep[eg.][]{Hoeijmakers2018, Hoeijmakers2019, Gibson2020, Seidel2021}, with a current popular theory to explain this suggesting that these heavier elements are dragged to these altitude by the escaping hydrogen \citep{Cubillos2020}. There are still many open questions to what the effects of magnetic field would be in the dynamics of heavier particles. For example, given that increasing the dipole strength increases Lyman-$\alpha$ absorption at line centre as more neutral hydrogen is trapped in the dead-zones, we can infer that this will also increase the line centre absorption of other heavier elements if they are well mixed with hydrogen in the escaping fluid. Likewise, the strong hydrogen outflow emerging from the poles could bring heavier elements to high altitudes. If this is the case, we could also expect  asymmetries in spectroscopic transits of heavier elements.

\subsection{Model Limitations}
Our simulations neglect the effects of charge exchange and radiation pressure. Radiation pressure from Lyman-$\alpha$ photons has been thought to accelerate neutral atoms to significantly blue-shifted velocities \citep{Bourrier2015, Schneiter2016}. However \citet{Debrecht2020} found that radiation pressure alone may not cause significant changes, with \citet{Carolina2021} finding that if the stellar wind already causes most of this acceleration, the contribution from radiation pressure will be minimal. Though the radiation pressure may have a small impact on the blue wing absorption it will not affect the mass loss rate, and so we do not expect the inclusion of radiation pressure to significantly alter our models. Charge exchange occurs when a stellar wind proton and a planetary neutral hydrogen atom exchange an electron at the boundary between the stellar wind and escaping atmosphere \citep{Shaikhislamov2016}. Though the number of neutrals remain the same, charge exchange will result in more high velocity neutral atoms thus increasing blue-shifted absorption \citep{2008Natur.451..970H, 2014A&A...562A.116K, Bourrier2016, Tremblin2013, Shaikhislamov2016}. Though these two processes are unlikely to affect the dynamics of these models, the transit line profiles may change \citep{2018MNRAS.475..605C}, with \citet{Esquivel2019, Odert2020} finding a combination of the two being the best fit for observations of HD209458b. One might expect that as a larger magnetosphere provides a larger interaction surface, that charge exchange will become more important for larger magnetic field strengths. However, because most of the low velocity neutrals are trapped in the dead-zones close the planet, and as the neutrals in the tail are already somewhat blue-shifted, we do not expect charge exchange to significantly alter the trends in absorption line profile that we see with increasing magnetic field strength, as it would only push already blue shifted material to higher velocities.

\section{Conclusions}
\label{sec:conclusions}

In this work, we have examined how magnetic fields affect the interaction between the stellar wind and escaping atmosphere. We use newly developed 3D self-consistent radiative magnetohydrodynamic models. Using the same magnetised stellar wind in each model, we vary the planetary dipole strength to examine how the planetary magnetic field affects the interaction with the stellar wind, the mass lost by the planet, and the observational signatures of this escape. \red{To the best of our knowledge this is the first 3D radiative model capable of simulating the inner most regions of the escaping atmosphere which includes: radiative heating \& cooling; cooling from collisions; the planetary magnetic field; magnetised stellar wind; Coriolis \& Centrifugal forces; and the force due to tidal gravity}.

We performed five simulations, varying the planetary dipole strength from 0 to 10G. We find that increasing the magnetic field strength of the planet greatly alters the structure of material in the magnetosphere. Just as was shown in other works \citep{Trammell2014, Owen2014, Khod2015, Arakcheev2017}, we found that the planetary magnetic field creates dead-zones, where the closed magnetic field lines around the planet trap material, reducing escape around the equator. \referee{This dead-zone grows with magnetic field strength, and is able to absorb more of the incident stellar radiation before it gets lower into the planet's atmosphere. As a result the $\tau=1$ surface is pushed to higher altitudes when the magnetic field strength is increased.} We also presented the novel finding of a double comet-like tail structure, one below and above the orbital plane, caused by the polar outflows. We found that the mass loss through the poles increases with magnetic field strength, causing an increase in the total atmospheric escape rate. This is due to the interaction with the stellar wind, which deflects dayside atmospheric escape through the poles into this double tail structure, placing further emphasis on the importance of considering the interaction of the escaping atmosphere with the stellar wind when investigating atmospheric escape. 

Using the results of our 3D simulations we investigated how changing the planetary magnetic field strength affects the observational signatures of atmospheric escape. We found an increase in line centre absorption with magnetic field strength, as more absorbing material is trapped in the growing dead-zones around the planet. The blue wing absorption initially decreases upon the introduction of the planetary field, as planetary material begins to be launched above and below the orbital plane, instead of being fully funnelled onto the orbital plane by the stellar wind, as seen in the 0G model. As the field strength continues to increase we see the blue wing absorption also increases. Similarly to the line centre absorption, the red wing absorption increases with magnetic field strength. We found that most of the red shifted material exists around the night-side orbital plane, as some material falls from the comet-like tails back towards the planet. 

Finally we investigated the asymmetry between positive and negative impact parameters during transit (i.e. if the planet transits in the northern or southern hemisphere of the star). With the exception of the 1G model (see appendix \ref{sec:appendix}), we find a growing asymmetry in the blue wing absorption with increasing magnetic field strength, with negative impact parameters leading to more absorption. This is caused by the interaction between the planetary dipole and radial stellar magnetic fields. Below the orbital plane, as the stellar wind is shocked back to sub-alfvénic velocities the stellar and planetary magnetic field lines reconnect (note that as the stellar wind is super-alfvénic before this interaction, this cannot affect the upstream stellar wind). Above the plane the opposite occurs, leading to more open planetary field lines. As a result we found that the planetary outflow is more extended above the plane than below for all magnetic field strengths. This causes asymmetry between positive and negative impact parameters, as depending on which comet-like tail is mostly covering the stellar disc during transit, different absorption profiles will be obtained. This work places importance not only on knowledge of the planetary magnetic field, but also on the geometry of its interaction with the  stellar wind's field when interpreting observational signatures of atmospheric escape.

\section*{Data Availability}
The data described in this article will be shared on reasonable request to the corresponding author.

\section*{Acknowledgements}
 \referee{We would like to thank the referee for their constructive review of this manuscript.} This project has received funding from the European Research Council (ERC) under the European Union's Horizon 2020 research and innovation programme (grant agreement No 817540, ASTROFLOW). The authors wish to acknowledge the SFI/HEA Irish Centre for High-End Computing (ICHEC) for the provision of computational facilities and support. This work used the BATS-R-US tools developed at the University of Michigan Center for Space Environment Modeling and made available through the NASA Community Coordinated Modeling Center. 
 
 \appendix
 
 \section{Quasi steady-state Solutions}
 \label{sec:appendix}
 The 1G model (and the 3G model, to a lesser extent) presented in this work does not settle to a fully steady-state solution. Instead we find a quasi steady-state where the resulting solution varies periodically with iteration number. Similar quasi steady-state solutions were found by \citet{McCann2019, Carolan2021}, with \citet{Christie2016} demonstrating that the amplitude of the variation in the solution decreases with increasing resolution in the grid.
 
\begin{figure}
    \centering
    \includegraphics[width=\columnwidth]{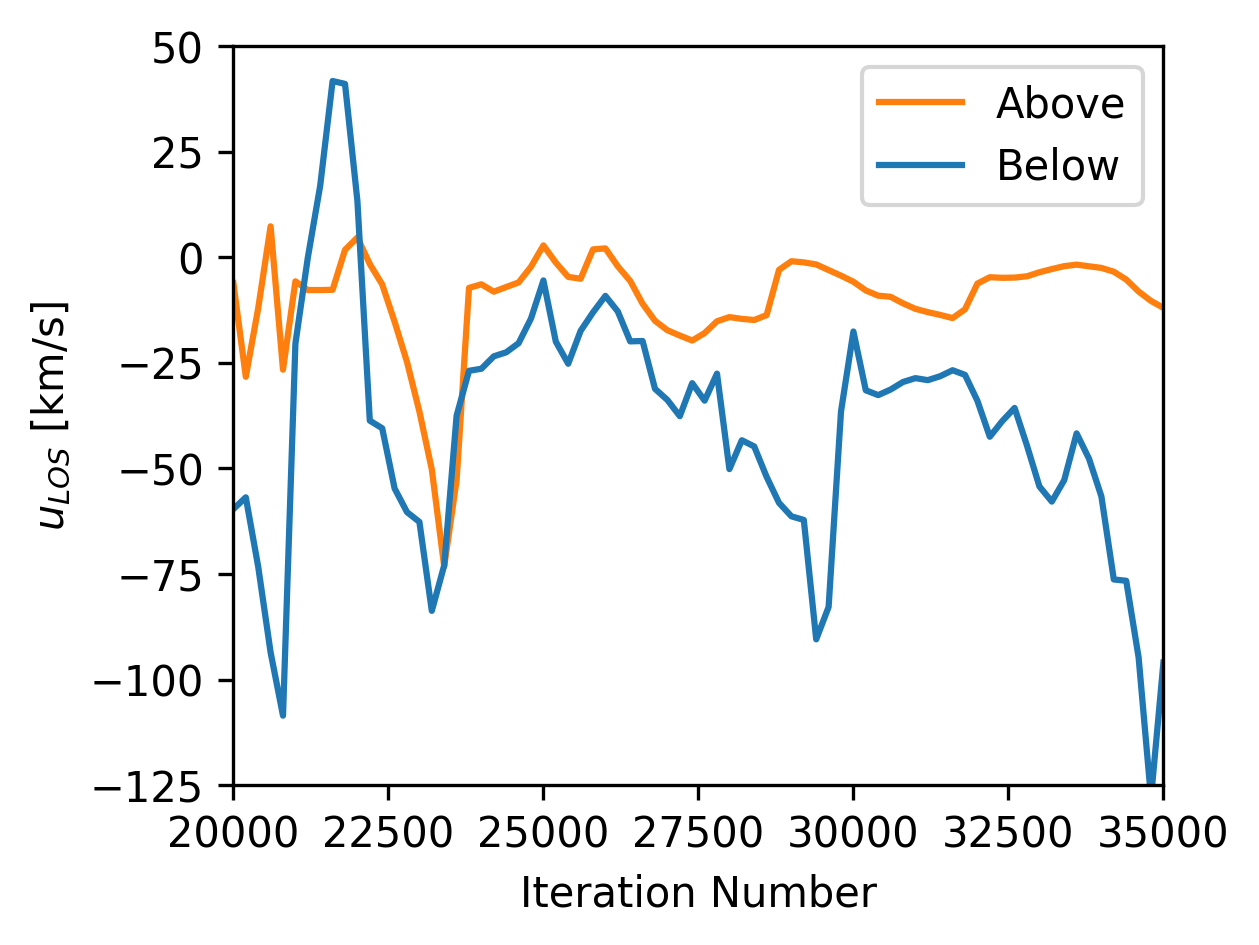}
    \caption{The quasi steady-state variation of the line of sight velocity in the 1G model. The two lines refer to points above and below the orbital plane, at X-Z coordinates [6, 3] and [6,-3] respectively. These points lie just within the magnetosonic surface in the left panel of the 1G model in figure \ref{fig:polar_plane}.}
    \label{fig:qss}
\end{figure}

 To illustrate the variation in this model, in figure \ref{fig:qss} we show how the line of sight velocity varies at two points in our grid, one above and below the orbital plane. Above the plane, we can see a small variation in the line of sight velocity after iteration 25000, with a period of  3000-4000 iterations, covering a range of velocities from 0 to -20 km/s. However below the plane we see a much larger variation, with the velocity ranging from -10 to -125 km/s, with a period of approximately 5000 iterations. This variation is responsible for the differences in northern/southern escape and absorption ins figures \ref{fig:mass_loss_rates} and \ref{fig:deltaF}.
 
The mass loss rates shown in figure \ref{fig:mass_loss_rates} for this model are calculated at the final iteration shown in figure \ref{fig:qss}. At this point, we can see that in the quasi steady-state solution, the magnitude of the velocity of material below the plane is much larger than that above the plane. As a result in this model we obtain a larger difference between mass lost through the north and south poles that what is found in other models.

Similarly, the effects of the quasi steady-state can be seen in the absorption of planetary material in figure \ref{fig:deltaF}. As the magnitude of the line of sight velocity below the plane it much larger at this stage in the solution, when the planet transits in the northern hemisphere of the star we see much more blue wing absorption than when it transits in the south. As we can see from figure \ref{fig:qss} this difference in absorption will vary depending on the stage of the solution examined (eg. at iteration 32000 the velocities above and below are much more similar), and so is not indicative of the overall trend we find in our models:\red{that the asymmetry between mid-transit absorption when transiting above and below the mid-disc will increase with the magnetic field strength of the planet.}
 
\bibliographystyle{mnras}
\bibliography{mybib}

\label{lastpage}
\end{document}